\documentclass{svproc}
\usepackage{url}
\usepackage{epsfig}

\begin{document}
\mainmatter              
\title{Optical Properties of Iron-Selenide Na$_{0.27}$K$_{0.27}$Rb$_{0.27}$Fe$_{1.7}$Se$_2$ Single Crystals}
\titlerunning{Optical Properties of Na$_{0.27}$K$_{0.27}$Rb$_{0.27}$Fe$_{1.7}$Se$_2$}  
%
\author{Andrei Muratov\inst{1} \and Yevgeny Rakhmanov\inst{1,2} \and
Andrei Shilov\inst{1} \and Igor Morozov\inst{2} \and Yurii Aleshchenko\inst{1}}
\authorrunning{Andrei Muratov et al.} 
\institute{P.N. Lebedev Physical Institute of RAS, 119991 Moscow, Russia,\\
\email{aleshchenkoya@lebedev.ru},\\
\and
Lomonosov Moscow State University,
Department of Chemistry, 119991 Moscow, Russia}

\maketitle              

\begin{abstract}
With infrared Fourier-transform spectroscopy and spectroscopic ellipsometry we investigated the in-plane reflectance of the superconducting Na$_{0.27}$K$_{0.27}$Rb$_{0.27}$Fe$_{1.7}$Se$_2$ single crystals with a critical temperature $T_c\approx 34.3$~K over a broad frequency range at temperatures of 4--300 K. The normal-state response of Na$_{0.27}$K$_{0.27}$Rb$_{0.27}$Fe$_{1.7}$Se$_2$ is analyzed by a Drude-Lorentz model with one Drude component. The temperature dependences of the plasma frequency, optical conductivity, scattering rate, and dc resistivity of the Drude component in the normal state are presented. We report a Fano-shaped mode at 1430~cm$^{-1}$ indicating a coupling of the discrete mode presumably related to some electronic transition to a two-magnon continuum as well as a low-frequency broad band assigned to the acoustic magnon branch of the antiferromagnetic superstructure. These features persist in the reflectance in both the normal and superconducting states suggesting a mesoscopic coexistence of superconductivity and magnetism in Na$_{0.27}$K$_{0.27}$Rb$_{0.27}$Fe$_{1.7}$Se$_2$.

\end{abstract}
\section{Introduction}
Among the iron-based superconductors (IBS), FeSe has a simplest layered structure with anti-PbO-type atomic arrangement~\cite{Chen 2014}. At ambient pressure, its critical temperature of superconducting  (SC) transition ($T_c$) is about 8~K~\cite{Hsu} but increases up to 37~K under a relatively small hydrostatic pressure of 6~GPa~\cite{Medvedev,Margadonna}. It has been shown~\cite{Wu1} that the microscopic effect of the applied pressure is to alter the separation of the Se atoms from the Fe planes, and its strong pressure dependence opens the possibility of raising $T_c$ by the introduction of internal chemical pressure. This possibility has been realized in a new class of SC $A_x$Fe$_2$Se$_2$ iron-selenides ($A$ = Na, K, Rb, Cs, K/Tl, Rb/Tl) with $T_c$ as high as 30--35~K~\cite{Guo,Mizuguchi,Krzton-Maziopa2011,Wang1,Ying1,Li,Wang2,Fang,Ying2}. These compounds possess a number of unique properties related to the inherent iron vacancy ordering resulting in the formation of a chiral $\sqrt {5}\times\sqrt {5}\times 1$ superstructure, which reduces the {\it I}4/{\it mmm} crystal symmetry to {\it I}4/{\it m}~\cite{Bacsa}. This $A_{0.8}$Fe$_{1.6}$Se$_2$ (245) phase is insulating and shows antiferromagnetic (AFM) long-range ordering up to very high N\'eel temperatures $T_{N}\sim 500$--600~K~\cite{Yuan2}. It coexists with the minority (10--20\%) $A_x$Fe$_2$Se$_2$ (122) vacancy-free conducting/semiconducting phase, which becomes superconductive at low temperatures. The minority phase is uniformly distributed within the volume of the crystal creating three-dimensional mesoscopic network of stripes~\cite{Krzton-Maziopa2016}. In contrast, it has been found recently using high spatial resolution techniques (magnetic force microscopy, Raman spectroscopy and scanning electron microscopy with energy dispersive x-ray spectroscopy)~\cite{Hou} that the stripe phase consists of relatively intact FeSe. Anyway, the resulting composition $A_x$Fe$_{2-y}$Se$_2$ is determined as an average over all the coexisting phases. Another striking feature of $A_x$Fe$_{2-y}$Se$_2$ selenides is the absence of a hole-like Fermi surface~\cite{Zhang,Wang4,Mou}. For this reason, the weak-coupling Fermi-surface-based mechanism will operate very differently in these materials compared with the iron pnictides~\cite{Qian,Wang5,Maier,Mati,Khodas,Wang6,Ghigo,Torsello}.

Optical spectroscopy is a versatile tool to study bulk electronic properties of superconductors both in the normal and SC state. Several infrared (IR) studies of $A_x$Fe$_{2-y}$Se$_2$ iron-selenides have been conducted~\cite {Chen,Yuan,Charnukha,Homes1,Ignatov,Wang3,Muratov}. However, the optical data on the properties of $A_x$Fe$_{2-y}$Se$_2$ are still scarce.

In this paper, we report an optical study of Na$_{0.27}$K$_{0.27}$Rb$_{0.27}$Fe$_{1.7}$Se$_2$ single crystals. According to previous studies~\cite{Roslova,Kuzmicheva2,Ilina}, the sodium doping facilitates the SC phase formation in $A_x$Fe$_{2-y}$Se$_2$ iron selenides. To the best of our knowledge, optical studies of Na$_{0.27}$K$_{0.27}$Rb$_{0.27}$Fe$_{1.7}$Se$_2$ have not been performed earlier.
\section{Experiment}
Large single crystals of Na$_{0.27}$K$_{0.27}$Rb$_{0.27}$Fe$_{1.7}$Se$_2$ up to 10~mm in dimension were grown using a self-flux technique. At the final stage, the preannealed and ground mixture of precursors was heated up to 1050$^\circ $C at a rate of 100$^\circ $C/hour in the sealed ampules and held at this temperature for 12 hours. Then, the ampoules were cooled down to 730$^\circ $C at a rate of 6$^\circ $C/hour and quenched in cold water. All manipulations with the preparation of precursors, storage and obtaining of samples for subsequent characterization, were performed in an argon-filled glove box with O$_2$ and H$_2$O content less than 0.1 ppm. The crystals were characterized by scanning electron microscopy equipped with the energy dispersive x-ray (EDX) spectroscopy, transmission electron microscopy, transport, and magnetic susceptibility measurements. Further details of the sample growth and characterization for the crystals from the same batch of samples can be found in~\cite{Rakhmanov}.

The reflectance $R(\omega )$ from the cleaved surface of Na$_{0.27}$K$_{0.27}$Rb$_{0.27}$Fe$_{1.7}$Se$_2$ was measured at a near normal angle of incidence on a Fourier transform IR spectrometer (Bruker IFS 125HR) for light polarized in the {\it ab} planes using an {\it in situ} aluminum evaporation technique~\cite{Homes93}. Data from 25 to 10000~cm$^{-1}$ were collected on a Konti Spectro A continuous-flow helium cryostat. To perform a Kramers-Kronig analysis, we extended the IR reflectance data to the visible and UV spectral ranges (4000--30 000~cm$^{-1}$) with a Woollam VASE spectral ellipsometer equipped with a high-vacuum Janis CRF 725V cryostat. In this frequency range, the complex optical conductivity was obtained directly from the measured ellipsometric coefficients. In the low-frequency region, we used a simple $\varepsilon _\infty $ + Drude approximation ($\varepsilon _\infty $ is the high-frequency dielectric constant), since the Hagen-Rubens approximation is valid for the samples with a pronounced metallic type of conductivity. All optical measurements were performed with a probing spot size $\sim 2.5$~mm at different temperatures, from 300~K down to 4~K.

\begin{figure}[!ht]
\includegraphics[width=8cm]{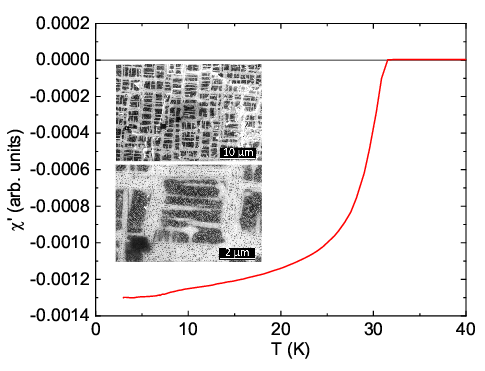}
\centering
\caption{(Color online) Temperature dependence of AC magnetic susceptibility for the quenched in cold water Na$_{0.27}$K$_{0.27}$Rb$_{0.27}$Fe$_{1.7}$Se$_2$ samples measured with a SQUID magnetometer MPMS XL7 in a magnetic field of 5~Oe at 337~Hz frequency. The inset shows secondary electron micrographs of Na$_{0.27}$K$_{0.27}$Rb$_{0.27}$Fe$_{1.7}$Se$_2$ taken at 5~keV with different magnifications (adapted from~\cite{Rakhmanov}).}
\end{figure}

It should be noted that all $A_x$Fe$_{2-y}$Se$_2$ selenides rapidly degrade in the open air and even in the presence of trace amounts of oxygen or water vapors, with $T_c$ turning to zero in several minutes. For this reason, the samples were transferred from the Ar filled ampoules to the optical cryostats with minimal delay. Freshly cleaved {\it ab}-plane sample surfaces were prepared at ambient conditions using Scotch tape just before pumping the cryostat.
\section{Results and discussion}
The obtained crystals have a mirror-like microtextured surface indicating the presence of different phases (see inset in Fig.~1). The secondary electron micrographs represent a network of perpendicular stripes with a thickness of about 0.5--1~$\mu $m. The dimensions of the rectangles bounded by these stripes are about 2--4~$\mu $m. There are tiny strips of smaller network within such rectangles directed mainly along one side of the rectangle. The dark regions within the rectangles consist of the AFM 245 phase; the tiny strips are the SC 122 phase, while each individual stripe of the main network contains more sodium but less potassium and iron representing some new phase [T.E. Kuzmicheva, private communication (unpublished)]. The EDX analysis yields the chemical formula of the samples as an average of all mentioned phases close to the nominal composition Na$_{0.27}$K$_{0.27}$Rb$_{0.27}$Fe$_{1.7}$Se$_2$.

The AC magnetic susceptibility $\chi '(T)$ of the samples is characterized by the onset critical temperature $T_{c\chi }\approx 31.5$~K (Fig.~1). The transport measurements show a sharp SC transition within 0.6~K with an onset at about 34.6~K. The $dR(T)/dT$ dependence represents a single maximum at $T_c\approx 34.3$~K attesting to high homogeneity of the SC properties (see Figs. 5c and 6a of~\cite{Rakhmanov} for the samples from the same batch as those used for this study). It is worth noting that the described sample preparation route provides good results only for the Na$_{0.27}$K$_{0.27}$Rb$_{0.27}$Fe$_{1.7}$Se$_2$ composition. The Na$_{0.4}$Rb$_{0.4}$Fe$_2$As$_2$ crystals prepared in the same way hardly demonstrate a SC transition, while the K$_{0.4}$Rb$_{0.4}$Fe$_2$As$_2$ samples have pure SC properties~\cite{Rakhmanov}.
\begin{figure}[!ht]
\includegraphics[width=8cm]{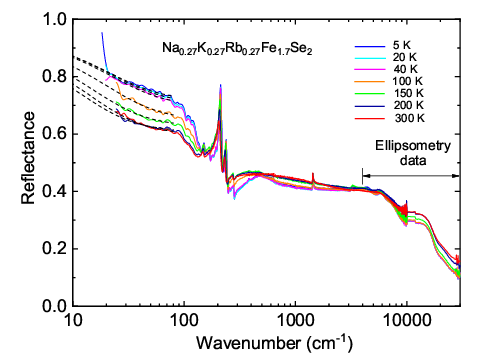}
\centering
\caption{(Color online) The absolute reflectance for Na$_{0.27}$K$_{0.27}$Rb$_{0.27}$Fe$_{1.7}$Se$_2$ single crystals for light polarized in the {\it ab} plane at several temperatures above and below $T_c$ in a broad frequency range. The dashed lines are fits with the simple $\varepsilon _\infty $ + Drude model.}
\end{figure}

Figure 2 shows the reflectance spectra of Na$_{0.27}$K$_{0.27}$Rb$_{0.27}$Fe$_{1.7}$Se$_2$ single crystals over a broad frequency range at selected temperatures above and below $T_c$. It can be seen that the overall reflectance is rather low, roughly close to the value of 0.4. This behavior is different from that of all other IBS as well as FeTe. The relatively low value of the reflectance is characteristic of a poor metal. Moreover, in contrast to other iron-pnictide compounds and, e.g., K$_{0.8}$Fe$_{1.7}$(Se$_{0.73}$S$_{0.27}$)$_2$~\cite{Muratov}, the high-frequency reflectance also shows noticeable temperature dependence. The far-IR $R(\omega )$ spectra are characterized by a free-carrier plasma edge at $\sim 150$~cm$^{-1}$ and numerous IR-active lattice vibrations. The latter as well as the electronic excitations at higher frequencies dominate in the spectra. For the tetragonal 122 structure of iron-pnictides having $I4/mmm$ crystal symmetry, only two in-plane IR-active $E_u$ phonon modes are allowed by the selection rules~\cite{Akrap}. More IR-active phonons support the reduced $I4/m$ crystal symmetry for which already 10 IR-active in-plane lattice modes are allowed~\cite{Bao}. The lattice modes narrow and shift to higher frequencies with decreasing temperature. The IR spectra also exhibit a narrow peak at 1430~cm$^{-1}$, which has no obvious change with temperature. We believe that this peak is due to electron-magnon interaction. It has an asymmetric Fano lineshape, indicating a coupling of the discrete mode presumably related to some electronic transition to a two-magnon continuum. The two-magnon peak has been observed earlier in high-frequency Raman spectra of iron-chalcogenide superconductors~\cite{Zhang2}. The Raman intensity of this peak decreased sharply on entering the SC state. On that ground, the authors asserted that superconductivity and magnetic order mutually excluded each other almost completely due to proximity effects within regions of microscopic phase separation. Our results contradict to~\cite{Zhang2} since the Fano-shaped peak is observed in IR reflectance in both the normal and SC states. Moreover, a microscopic coexistence of superconductivity and a strong magnetic order in iron selenides has been confirmed by $\mu $SR~\cite {Shermadini} and neutron diffraction~\cite {Ye,Bao} studies. The line shape of this peak will be analyzed below. The low-frequency reflectance starts to increase below 200~K showing a week metallic response. At temperatures approaching $T_c$, the reflectance below $\sim 150$~cm$^{-1}$ exhibits steep increase with the formation of a relatively sharp plasma edge. However it does not reach unity in the measurement range which may be explained by the low volume fraction of the SC phase.
\begin{figure}[!ht]
\includegraphics[width=8cm]{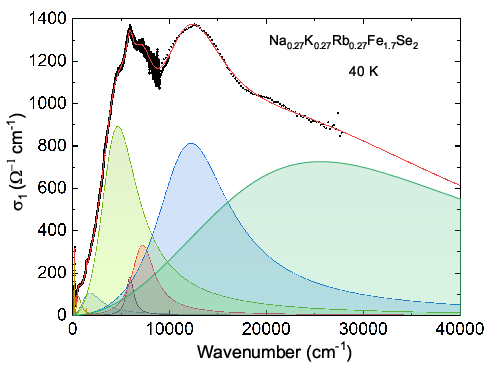}
\centering
\caption{(Color online) The real part of the optical conductivity of Na$_{0.27}$K$_{0.27}$Rb$_{0.27}$Fe$_{1.7}$Se$_2$ at 40~K over a broad frequency range (dots) together with the Lorentz contributions revealed by the Drude-Lorentz analysis. The red solid curve is a total contour.}
\end{figure}

For more convenient analysis, we converted our reflectance $R(\omega )$ into the real part of the optical conductivity $\sigma _1(\omega )$ by the Kramers-Kronig transformation. The optical response of Na$_{0.27}$K$_{0.27}$Rb$_{0.27}$Fe$_{1.7}$Se$_2$ single crystals in the normal state was modeled using a Drude-Lorentz model for the complex dielectric function with one Drude component and a set of Lorentz contributions representing the lattice vibrations and interband transitions. In this case, the complex dielectric function $\tilde\varepsilon (\omega)=\varepsilon _1(\omega)+i\varepsilon _2(\omega)$ can be written in the form
$$
\tilde\varepsilon (\omega)=\varepsilon _{\infty }-\frac{\omega _{D,pl}^2}{\omega (\omega +i\gamma _{D})}+\sum _i\frac{\omega _{i,pl}^2}{\omega _i^2-\omega ^2-i\omega\gamma _i},
$$
where $\omega _{D,pl}$ is the Drude plasma frequency, $\gamma _D$ is the scattering rate of free (Drude) charge carriers, $\omega _{i,pl}$, $\omega _i$ and $\gamma _i$ denote the plasma frequency, the center frequency, and the damping of the {\it i}th Lorentz component, respectively. The optical conductivity can be related to the dielectric function as $\tilde\sigma (\omega )=\sigma _1(\omega)+i\sigma _2(\omega )=i\omega [\varepsilon _\infty -\tilde\varepsilon (\omega )]/4\pi $.

As an example, in Fig. 3 we show $\sigma _1(\omega )$ spectra of Na$_{0.27}$K$_{0.27}$Rb$_{0.27}$Fe$_{1.7}$Se$_2$ at 40~K, focused on a high-frequency range, together with the Lorentz contributions revealed by the Drude-Lorentz analysis. One can see a number of broad bands superimposed on the wide hump. The broad bands with the maxima at $\sim 760$, 4600, 5900, 7100, and 13000~cm$^{-1}$ have been observed earlier in optical studies of the K- and Rb-doped iron selenides~\cite {Chen,Yuan,Charnukha,Homes1} and were ascribed to the spin-controlled interband transitions. In our previous paper~\cite {Muratov}, we also observed broad bands at $\sim 1000$, 4400, 5900, 6570, and 12800~cm$^{-1}$ for K$_{0.8}$Fe$_{1.7}$(Se$_{0.73}$S$_{0.27}$)$_2$. We believe that these bands as well as the bands at 600, 1790, 4670, 5920, 7220, and 12260~cm$^{-1}$ observed in the present paper (see Fig.~3) also have a similar origin.
\begin{figure}[!ht]
\includegraphics[width=8cm]{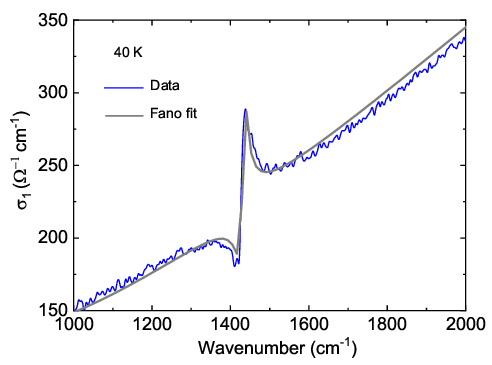}
\centering
\caption{(Color online) Representative fit (thick line) of Fano lineshape to IR reflectance data at 40~K.}
\end{figure}

As we have already mentioned, the asymmetric Fano line shape of the mode at 1430~cm$^{-1}$ in the reflectance spectrum attracts attention. In the spectrum of the real part of the optical conductivity, we have modeled this mode as a Fano resonance~\cite{Damascelli},
$$
\sigma _{Fano}(\omega )=i\sigma _0\frac {(q-i)^2}{i+x},
$$
with $x =(\omega ^2-\omega _0^2)/\gamma\omega $, $\sigma _0=\omega _{pl,Fano}/4\pi\gamma q^2$. Here $\omega _0$, $\omega _{pl,Fano}$, $\gamma $, and $q$ are the resonance energy, the plasma frequency, the linewidth, and the Fano asymmetry parameter of the modes, respectively. The Fano parameter $q$ determines the asymmetry of the resonance and is equal to 1.58 in the case of $T=40$~K (see Fig.~4). The inverse of the Fano asymmetry parameter, $1/q=0.63$, represents the relatively high strength of the electron-magnon coupling and quantifies the degree of the asymmetry. When $|1/q|\ll 1$, the Lorentz line shape is recovered.
\begin{figure}[!ht]
\includegraphics[width=8cm]{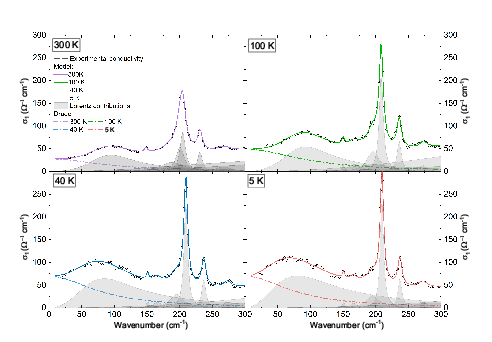}
\centering
\caption{(Color online) The expanded plots of the $\sigma _1(\omega )$ spectra below 300~cm$^{-1}$ taken at 300~K, 100~K, 40~K, and 5~K together with the results of Drude-Lorentz fit to the conductivity of Na$_{0.27}$K$_{0.27}$Rb$_{0.27}$Fe$_{1.7}$Se$_2$. The Lorentz contributions are shown in gray.}
\end{figure}

Figure 5 represents the optical conductivity spectra of the Na$_{0.27}$K$_{0.27}$Rb$_{0.27}$Fe$_{1.7}$Se$_2$ single crystals at selected temperatures in the expanded low-frequency region as well as the results of a nonlinear least-squares fit to the $\sigma _1(\omega )$ using a single Drude contribution and a series of Lorentzians reproducing the phonon peaks. In this figure, the dots are obtained via Kramers-Kronig analysis, the solid lines through the data are fits with the Drude-Lorentz model, the Lorentz contributions are shown in gray, and the dash-dotted lines are the Drude contributions. The physical parameters obtained by fitting the in-plane IR-active lattice modes in the $\sigma _1(\omega )$ spectra of Na$_{0.27}$K$_{0.27}$Rb$_{0.27}$Fe$_{1.7}$Se$_2$ at 40~K are listed in Table~1. There is a very weak Drude response at 300~K in the conductivity spectrum (Fig.~5), which gradually rises with decreasing temperature. We also notice a broad band in the low frequency region. Its intensity and width increase with decreasing temperature while its maximum shifts to lower energies. It cannot be ascribed to the phonon mode because of its broad width. We suppose that this band is related to the acoustic magnon branch of the antiferromagnetic $\sqrt {5}\times\sqrt {5}\times 1$ superstructure, which extends to 70~meV in Rb$_{0.89}$Fe$_{1.58}$Se$_2$~\cite {Wang7}. The origin of this band deserves further study. Meanwhile, the presence of this band as well as the strong phonon modes obscure the gapping of the SC phase.
{\sloppy

}

\begin{table*}[!ht]
\footnotesize
\centering
\begin{tabular}{|c|c|c|}
  \hline
$\omega _i$ & $\gamma _i$ & $\omega _{i,pl}$ \\
  \hline
151.0 & 4.25 & 61.1 \\
  \hline
165.9 & 16.7 & 111.4 \\
  \hline
195.3 & 23.7 & 207.0 \\
  \hline
209.4 & 6.7 & 307.5 \\
  \hline
236.9 & 7.5 & 167.9 \\
  \hline
263.5 & 7.0 & 54.4 \\
  \hline
272.8 & 13.8 & 104.3 \\
  \hline
\end{tabular}
\caption{Physical parameters obtained by fitting the in-plane IR-active lattice modes in the optical conductivity of Na$_{0.27}$K$_{0.27}$Rb$_{0.27}$Fe$_{1.7}$Se$_2$ at 40~K, where $\omega _i$, $\gamma _i$, and $\omega _{i,pl}$ are the frequency, width, and plasma frequency of the $i$th mode. All units are in cm$^{-1}$.}
\end{table*}
\begin{figure}[!ht]
\includegraphics[width=8cm]{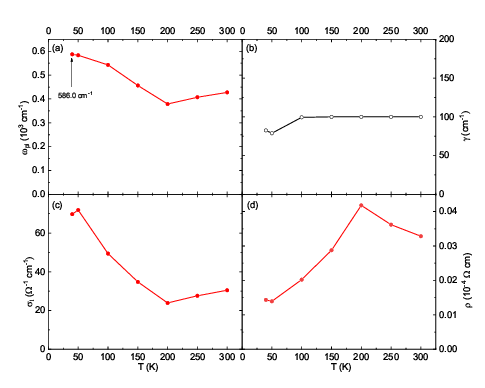}
\centering
\caption{(Color online) The model parameters $\omega _{pl}$ (a), $\gamma $ (b), $\sigma _1$ (c), and $\rho $ (d) of the Na$_{0.27}$K$_{0.27}$Rb$_{0.27}$Fe$_{1.7}$Se$_2$ single crystals in the normal state as a function of temperature.}
\end{figure}

The temperature dependences of the normal-state model parameters, such as the plasma frequency $\omega _{pl}$, the static scattering rate $\gamma $, the real part of the optical conductivity $\sigma _1$, and the dc resistivity $\rho $ of the Na$_{0.27}$K$_{0.27}$Rb$_{0.27}$Fe$_{1.7}$Se$_2$ single crystals are depicted in Fig.~6. The proximity to a number of strong phonon modes (see Fig.~5 and Table~1) brings uncertainty in the parameters provided that $\gamma $ is not fixed. To overcome this problem, we fix the static scattering rate in the region of 100--300~K (Fig.~6b). The plasma frequency $\omega _{pl}=586$~cm$^{-1}$ determined from the Drude-Lorentz fit to the optical conductivity at 40~K is more than an order of magnitude smaller than that in comparable iron-arsenide~\cite {Tu} or iron-chalcogenide materials~\cite {Homes2}. The Drude resistivity (Fig.~6d) shows a similar temperature dependence as the resistance of the quenched Na$_{0.27}$K$_{0.27}$Rb$_{0.27}$Fe$_{1.7}$Se$_2$ samples from the same batch determined with Van der Pauw method~\cite {Rakhmanov} and is typical for the $A_x$Fe$_{2-y}$Se$_2$ family. Just above $T_c$, it shows a metallic-like growth, forms a hump at $T\approx 200$~K, then rapidly decreases at higher temperatures in a semiconductor-like way. Such behavior could originate from the natural phase separation in the $A_x$Fe$_{2-y}$Se$_2$ selenides.

\section{Conclusions}

In conclusion, we have studied the optical properties of the superconducting Na$_{0.27}$K$_{0.27}$Rb$_{0.27}$Fe$_{1.7}$Se$_2$ single crystals in the broad spectral range from 25 to 30 000 cm$^{-1}$ at temperatures 4--300~K. A one-Drude-Lorentz model was found to describe successfully the optical response of Na$_{0.27}$K$_{0.27}$Rb$_{0.27}$Fe$_{1.7}$Se$_2$ in the normal state. The temperature dependences of the plasma frequency, optical conductivity, scattering rate, and dc resistivity of the Drude term in the normal state are presented. A coupling of the discrete mode presumably related to some electronic transition to a two-magnon continuum manifests itself as a Fano-shaped mode at 1430~cm$^{-1}$, which persists in the IR spectra of Na$_{0.27}$K$_{0.27}$Rb$_{0.27}$Fe$_{1.7}$Se$_2$ in both the normal and SC states. We relate a low-frequency broad band also detected in both states to the acoustic magnon branch of the antiferromagnetic 245 superstructure. Our results support a mesoscopic coexistence of superconductivity and magnetism in Na$_{0.27}$K$_{0.27}$Rb$_{0.27}$Fe$_{1.7}$Se$_2$.
\paragraph{Acknowledgements}
The authors would like to thank Vladimir Men'shov and Giovanni A. Ummarino for helpful comments on the manuscript. The research was supported by Russian Science Foundation grant No. 22-72-10082-P. The measurements were done using research equipment of the Shared Facilities Center at LPI.
%
%

\end{document}